*Article*

# Prediction of Food Production Using Machine Learning Algorithms of Multilayer Perceptron and ANFIS

Saeed Nosratabadi[1], Sina Ardabili[2], Zoltan Lakner[3], Csaba Mako[4,*], Amir Mosavi[5,6,*]

[1] Doctoral School of Economic and Regional Sciences, Hungarian University of Agriculture and Life Sciences, 2100 Godollo, Hungary; saeed.nosratabadi@phd.uni-mate.hu (S.N.)
[2] School of Economics and Business, Norwegian University of Life Sciences, 1430 Ås, Norway; s.ardabili@ieee.org (S.A.) and amir.mosavi@mailbox.tu-dresden.de (A.M.)
[3] Institute of Economic Sciences, Hungarian University of Agriculture and Life Sciences, 2100 Godollo, Hungary; Lakner.Zoltan.Karoly@uni-mate.hu (Z.L)
[4] Institute of Information Society, University of Public Service, 1083 Budapest, Hungary; mako.csaba@tk.mta.hu (C.M.)
[5] Faculty of Civil Engineering, Technische Universitat Dresden, 01069 Dresden, Germany
[6] John von Neumann Faculty of Informatics, Obuda University, 1034 Budapest, Hungary
*Correspondence: mako.csaba@tk.mta.hu, amir.mosavi@mailbox.tu-dresden.de

**Abstract:** Advancing models for accurate estimation of food production is essential for policymaking and managing national plans of action for food security. This research proposes two machine learning models for the prediction of food production. The adaptive network-based fuzzy inference system (ANFIS) and multilayer perceptron (MLP) methods are used to advance the prediction models. In the present study, two variables of livestock production and agricultural production were considered as the source of food production. Three variables were used to evaluate livestock production, namely livestock yield, live animals, and animal slaughtered, and two variables were used to assess agricultural production, namely agricultural production yields and losses. Iran was selected as the case study of the current study. Therefore, time-series data related to livestock and agricultural productions in Iran from 1961 to 2017 have been collected from the FAOSTAT database. First, 70% of this data was used to train ANFIS and MLP, and the remaining 30% of the data was used to test the models. The results disclosed that the ANFIS model with Generalized bell-shaped (Gbell) built-in membership functions has the lowest error level in predicting food production. The findings of this study provide a suitable tool for policymakers who can use this model and predict the future of food production to provide a proper plan for the future of food security and food supply for the next generations.

## 1. Introduction

Climate change, natural hazards, drought, uncertainty in recourses, and population growth are increasingly threatening the food security of the global nations [1]. It is estimated that the world's population will exceed 9.7 billion by 2050, which will encourage worldwide hunger and food insecurity [2]. In general, there are two means of the food supply, i.e., domestic production and imports [3]. Awareness of a region's potential for producing food provides the foundation for developing informed policies for food security. Thus, advancing accurate prediction models is considered essential for food governance and business models [4]. Reliable food prediction models can be used by policymakers to reconsider the annual food import volumes and prices [5]. Furthermore, insight into the food production value to better manage the poverty and support vulnerable groups exposed to food insecurity [6]. Conventional time series and mathematical models had been often used to project food production [7]. Advanced data-driven methods based on artificial intelligence and machine learning have recently shown promising results in providing accurate prediction models. The research for the advancement of reliable artificial intelligence and machine learning methods to be used in a higher level of policymaking is still in the early stage [8-10].

    A review of the literature for studies that predicted agricultural and livestock production, as the essential representatives of food production, shows that the available studies at the micro-level often focus on a specific crop or individual livestock. For instance, Nosratabadi et al. [7], Pantazi et al. [8], and Sengupta and Lee [9], used machine learning techniques to develop models for crop yield prediction. Nosratabadi et al. [7] develop gray wolf optimizer of neural networks (GWO-ANN), a hybrid machine learning model, to predict the yield of wheat crops in Iran and they also state that this model has a lower error rate and higher predictive accuracy (with R= 0.48 and root mean square error (RMSE)=3.19) compared to other models. Pantazi et al. [8] designed a supervised Kohonen networks (SNK) model to predict wheat yield. They report that the accuracy of their model in the prediction of wheat yield was 81.65%. Sengupta and Lee [9] using a support vector machine (SVM) tried to identify the number of immature green citrus and they report that the accuracy of their model was 80.4%. In addition, Morales et al. [10], Alonso, Villa, and Bahamonde [11], and Alonso, Castañón, and Bahamonde [12], for example, have employed machine learning techniques to design models for livestock production. Morales et al. [10] develop an SVM model for the early detection of problems in the production curves of hens' eggs. They claim that the accuracy of their mode has been equal to 98%. Alonso et al. [11] developed an SVM model to forecast cattle weight trajectories with only one or a few weights. And they report that the level of error metrics of mean absolute percentage error (MAPE) for their model were between 3.9 to 9.3 for different datasets. Alonso et al. [12] develop an SVM/ support vector regression (SVR) to estimate the beef cattle' carcass weight 150 days before slaughter. They used MAPE to test the accuracy of their model and they report that the Average MAPE of their model was 4.27%. Although research has used advanced machine learning tools to predict agricultural and livestock production, the focus of the research has been on a specific product or livestock, and developed models are not designed to forecast different production at the macro level of a country. To address this gap in the literature, the present study intends to develop a model for predicting food production at the macro level of a country using machine learning models.

Since there is ample evidence that agriculture in Iran is facing many problems due to a lack of water resources (e.g., Karandish et al. [13] and Qasemipour and Abbasi [14]), with successive droughts (e.g., Paymard et al. [15]) and poor water management (e.g., Raeisi et al. [16] and Akhoundi and Nazif [17]) cited as reasons for Iran's lack of water. Such problems have hampered food security at the macro level in Iran. On the other hand, Iran, with 79 million in 2015 [18], is one of the most populous countries in the world and is expected to have positive population growth in Iran in the future [18]. There are plenty of studies that explain that some Iranian households are exposed to food insecurity for reasons such as low levels of education and low levels of income (e.g., Ekhlaspour et al. [19], Esfarjani et al. [20], Fathi Beyranvand et al. [21], Najafi Alamdarlo et al. [22]). Therefore, in the present study, Iran was selected as a case study, and the time-series data of agricultural and livestock products related to Iran were used to develop and test the research model.

In the literature, there are advanced and accurate methods for predicting future trends using past data. Artificial intelligence models have the ability to learn from data and can predict non-linear phenomena with very high accuracy based on existing data. There is ample evidence that neural networks, as one of the tools of artificial intelligence, have a very high performance in predicting time series data. For example, Tealeb [23] conducts a review study detailing the articles that used artificial neural networks (ANN) to predict time series data and shows that the results of ANN are promising in predicting time series data. On the other hand, Tealab, Hefny, and Badr [24] debate that it is better to use advanced and hybrid ANN models in predicting non-linear time series data. Adaptive network-based fuzzy inference system (ANFIS) is a hybrid ANN that is combined with fuzzy systems that can be applied for the time-series data. Hence, the main objective of the current study is to compare the predictive performance of multilayer perceptron (MLP), a type of ANN, and ANFIS in the prediction of the future of agricultural and livestock production in Iran to select the most accurate model. The output of the present study provides policymakers with a comprehensive picture of the future food supply in Iran. Information on predicting indigenous food production provides knowledge to macro-decision makers to design appropriate policies for food security and provide adequate food for future generations. The research has been designed based on a comparative analysis of MLP and ANFIS. Our study investigates the model performance of neural networks and neuro-fuzzy.

The structure of the manuscript is represented as follows. First, the data, data source, and the data collection process are elaborated. The machine learning methods used in this paper are then described in detail. After that, the results of comparing MLP and ANFIS are presented. In the next stage, the most accurate model for predicting food production based on the results of accuracy metrics is presented.

## 2. Research Background

*2.1 Food Security in Iran*

Iran is one of the countries exposed to drought [15] as climate change and inadequate agricultural irrigation systems are among the main reasons mentioned in the literature for the problem of drought in Iran [25]. Drought is a serious threat to food security and has created many challenges for food supply in Iran. Iran is a vast country with diverse climatic conditions that have led to the cultivation of various agricultural products in different parts of the country. Drought and rising population growth, nonetheless, have jeopardized food supply and food security in the country. Qasemipour and Abbasi [14] believe that intensive agricultural practices in Iran led to water scarcity of 206%. Of course, research solutions have been proposed to address water management in order to increase food security and improve food production in Iran. Raeisi et al. [16], for example, consider greenhouses as an alternative to traditional farming because of better water management and higher crop yields. On the other hand, Akhoundi and Nazif [15] propose a model by which wastewater is used to irrigate agricultural fields instead of using natural water. Besides, Esfahani et al. [26] introduce a more creative model to deal with water scarcity in Iran. They consider overseas cultivation as a solution to contribute to food security in Iran.

*2.2. Application of Data Science in Food and Agriculture*

Many researchers have used data science to solve research problems related to food and agriculture. Since machine learning and deep learning models have the ability to analyze big data, find trends, and make accurate predictions, they have become highly useful tools for researchers [27]. Sengupta and Lee [28] and Su, Xu, and Yan [29], for instance, have used the SVM model and Ali et al. [30] has used the ANFIS model to predict crop yield. The use of learning machine to detect diseases is one of the other applications of machine learning in agriculture. For example, Chung et al. [31] and Ebrahimi et al. [32] used the SVM model to detect diseases in rice and strawberry crops, respectively. The use of ANN models to detect wheat diseases has been very common. So that Moshou et al. [33] has used the ANN/MLP model, Moshou et al. [34] employ the ANN/SOM model to detect wheat diseases. There are also studies that have used machine learning models to detect weeds. For example, Pantazi et al. [35] and Pantazi, Moshou, and Bravo [36] use an ANN model to detect weeds. Water management and soil management are other applications that have used machine learning models to improve agricultural production. For example, Feng et al. [37] and Patil and Deka [38] use the ANN model to estimate evapotranspiration. Estimation of soil temperature and humidity are also among the applications of machine learning models for soil management. In addition, the use of machine learning models to solve problems related to livestock management has become trendy. Craninx et al. [39], for example, has used the ANN model to forecast rumen fermentation pattern from milk fatty acids in cattle. Alonso, Villa, and Bahamonde [40] uses the SVM model to estimate the weight of cattle at different stages of growth with the least number of weights. Alonso, Castañón, and Bahamonde [41] also used the SVM model to predict carcass weight for beef cattle 150 days before slaughter.

Researchers have also used machine learning models in the food industry. The main applications of machine learning and deep learning in food are to estimate the quality of food. For example, Liu et al. [42] combined stacked sparse autoencoder (SSAE) with CNN to develop a model that detect the quality of vegetables. In addition, Rodriguez et al. [43] and Azizah et al. [44] use CNN to study the quality of fruits. There are studies that evaluate the quality of meat and aquatic products using deep learning models [45-46]. Using machine learning models to study food contaminations is another example of using machine learning in the food industry [47-48].

**3. Materials and Methods**

*3.1. Data*

The aim of this study is to develop a model for predicting food production for the next decade in Iran. In the present study, two sub-variables of agricultural production and Livestock production have been considered to evaluate food production. Three variables, livestock yield, live animals, and animal slaughtered, are used to measure livestock production. This study has also considered two variables, agricultural production yields, and losses, to evaluate the agricultural production. Figure 1 represents the model of the study. In this study, the production of barley, beans, dates, maize, millet, potatoes, rice, soybeans, wheat, rye, and olives is considered as agricultural production in Iran. According to this model, agricultural production yields and losses of the aforementioned products are evaluated as two input variables of agricultural production. Since the losses refer to the loss of productions the respective arrow is drawn outward. For the livestock production, the data related to the live animals such as beehives, buffalo, camel, cattle, chicken, duck, geese, goat, pig, sheep, end turkey and the data related to indigenous meat of buffalo, camel, cattle, chicken, duck, geese, goat, pig, sheep, and turkey and the data related to milk of buffalo milk, cow, goat, and sheep are collected. These data are collected from the FAO database, i.e., FAOSTAT, that can be accessed on http://www.fao.org/faostat/en/#data. The collected data covers the period of 1961-2017.

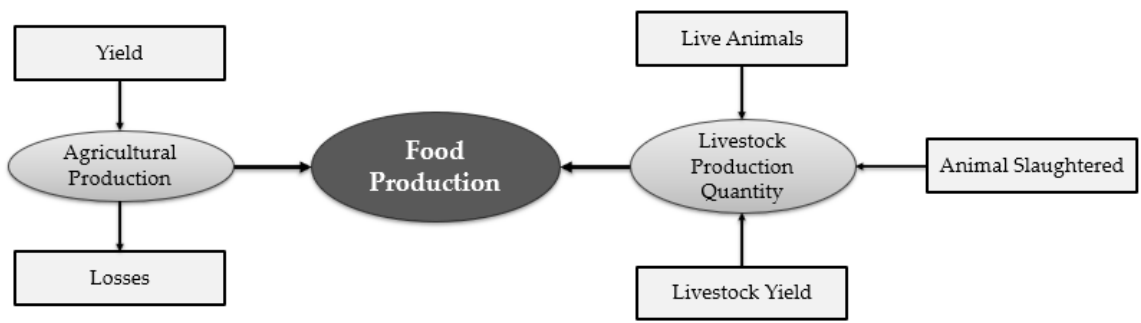

Figure 1. The proposed model of the study for indigenous food production in Iran

Figure 1 shows that the indigenous livestock production quantity and the indigenous agricultural production in Iran are considered as the country's potential food production for this country. Two variables of yield and losses were used to evaluate and measure agricultural production, and three variables of live animals, livestock yield, and slaughtered animals were used to measure livestock production quantity.

*3.2. Methods*

For predicting the future trends of food production in Iran, two models of MLP and ANFIS are applied in the collected data, and the predictive performance of the models are compared based on the accuracy metrics. We trained the proposed models by minimizing a regularized loss function on the training set and evaluated the models by comparing the accuracy metrics on the test set.

3.2.1. MLP

Multilayer Perceptron (MLP) is a type of neural network that has a supervised learning technique using the back-propagation method. Figure 2 shows that MLP benefits from a three-layer structure, including the input layer, hidden layer/s, and output layer/s, in which each neuron is connected to all the neurons in the next layer. It is frequently reported that MLP has a great function in non-linear problems [49-50].

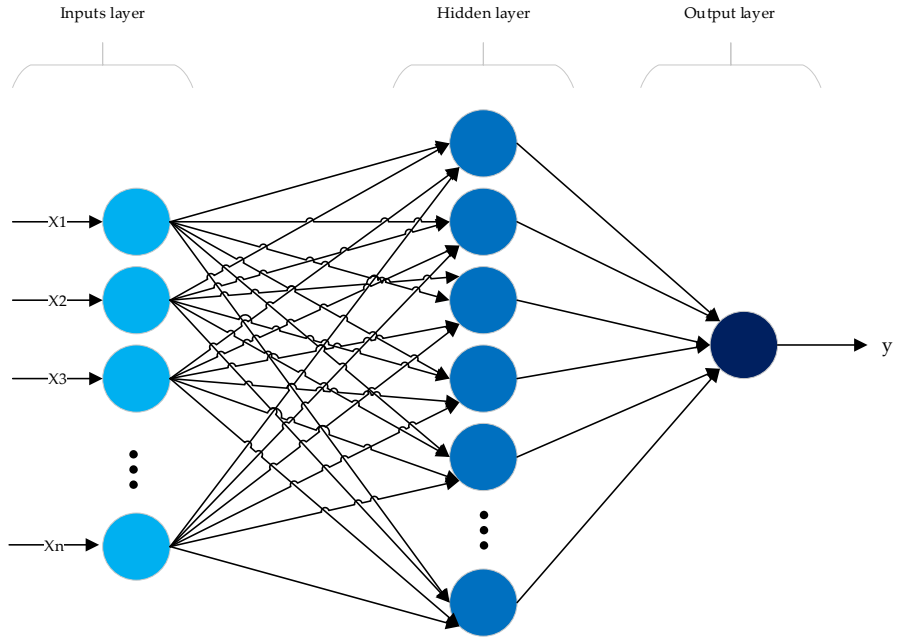

Figure 2. The architecture of the Multilayer Perceptron neural networks.

Equation (1) shows how the output of input variables, bias values, and input values are calculated:

$$S_i = \sum_{i=1}^{n} w_{ij} I_i + \beta_i$$
(1)

Where I represent the input layer, $I_i$ is the input variable i, n shows the total number of inputs, $\beta_j$ is a bias value, $\omega_{ij}$ is the weight of connections in j level. The sigmoid function is mostly used as the activation functions in MLP and it can be calculated through Equation (2):

$$f_j = \frac{1}{1+e^{-S_j}}$$
(2)

where, S is the activation function. Therefore, the ultimate output neuron j can be measured Equation (3):

$$y_i = f_i(\sum_{i=1}^{n} w_{ij} I_i + \beta_i) \quad (3)$$

where, y presents the output value of the MLP method which needs to be compared by the target values for calculating the model performance. MLP was trained by 70% of total data as a training dataset which has been sorted randomly by the model. The training was performed by different sets of the neuron numbers in the hidden layer for finding the best architecture for the predictor model from 10 to 18 by interval 4. The activation function was selected to be the Tanh(x) due to its higher performance compared with other activation functions.

### 3.2.2. ANFIS

The adaptive network-based fuzzy inference system is a hybrid neural network in which a fuzzy logic (FL) is embedded to the artificial neural network (ANN) architecture to identify the optimal distribution of membership functions [51]. The inference system of ANFIS consists of five layers in which the input of each layer is the output of the previous layer. This method applies fuzzy if-then rules of Sugeno, and if an ANFIS model has two inputs (x, y) and one output ($f_i$), for example, the two rules for a first-order two-rule are:

- Rule 1: if x is $A_1$ and y is $B_1$ then z is $f_1(x, y)$
- Rule 2: if x is $A_2$ and y is $B_2$ then z is $f_2(x, y)$

Where x and y are the ANFIS inputs, A and B are the fuzzy sets, $f_i(x, y)$ is the outputs of the first-order Sugeno fuzzy. The architecture of an ANFIS model constitutes adaptive nodes and fixed nodes (see Figure 3). The first layer of the model includes adaptive nodes that can be calculated through Equations 4, 5, and 6.

$$O_{1,i} - \mu_{A_i}(x) \; for \; i = 1,2 \quad (4)$$

$$O_{1,i} - \mu_{B_i}(y) \; for \; i = 1,2 \quad (5)$$

$$\mu(x) = \frac{1}{1+(\frac{x-c_i}{a_i})^{2b_i}} \quad (6)$$

Where x and y are the inputs, A and B are the linguistic labels, $\mu(x)$ and $\mu(y)$ are membership functions that take values between 0 and 1, and $a_i$, $b_i$, and $c_i$ are the parameter sets.

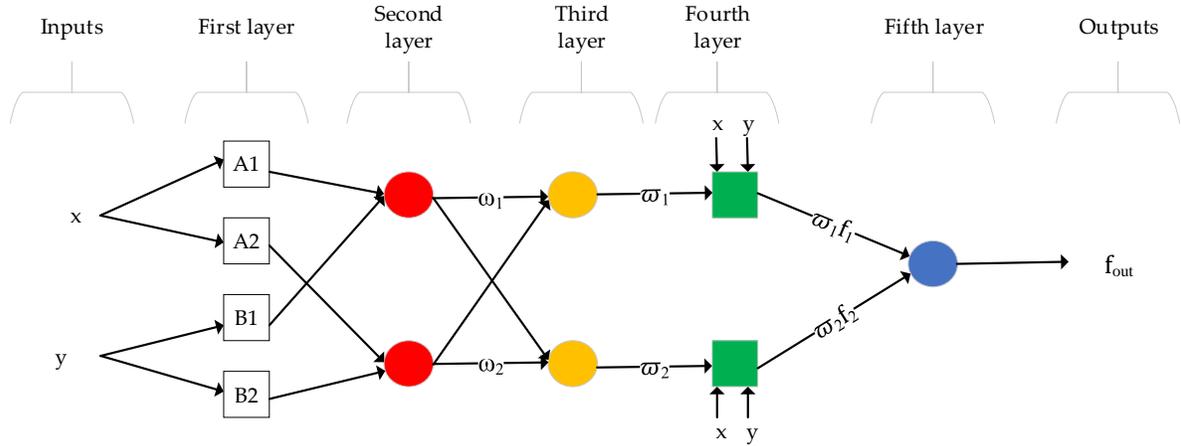

**Figure 3.** The architecture of Adaptive Network-Based Fuzzy Inference System.

The second layer, which is shown in red circles in Figure 3, is a fixed node and can be calculated through Equation 7. It is worth mentioning that $\omega_i$ is the firing strength of a rule.

$$O_{2,i} = w_i = \mu_{Ai}(x) \cdot \mu_{Bi}(y), \quad \text{for i = 1,2} \tag{7}$$

$O_{2,i}$ as the output of the second layer enters to the third layer. The third layer, which is presented in yellow circles in Figure 3, is also a fixed node. Its main goal is to normalize the firing strength by using Equation 8.

$$O_{3,i} = \overline{w}_i = \frac{w_i}{\sum w_i} = \frac{w_i}{w_1 + w_2}, \quad \text{for i = 1,2} \tag{8}$$

The fourth layer is an adaptive node as well and depicted as green squares. Equation 9 is used to measure the fourth layer.

$$O_{4,i} = \overline{w}_i \cdot f_i, \quad \text{for i = 1,2} \tag{9}$$

- Rule 1: if x is $A_1$ and y is $B_1$ then $f_1 = p_1x + q_1y + r_1$
- Rule 2: if x is $A_2$ and y is $B_2$ then $f_2 = p_2x + q_2y + r_2$

Where $p_i$, $q_i$, and $r_i$ are the parameters sets.

The fifth layer is also a fixed node presented in the form of a blue circle in Figure 3 and can be calculated through Equation 10.

$$O_{5,i} = f_{out} = \sum \overline{w}_i \cdot f_i = Overal\_output, \quad \text{for i = 1,2} \tag{10}$$

The final output of an ANFIS structure, which is shown as $F_{out}$ in Figure 3, can be calculated through Equation 11.

$$cf_{out} = \overline{w}_1 f_1 + \overline{w}_2 f_2 = \frac{w_1}{w_1 + w_2} f_1 + \frac{w_2}{w_1 + w_2} f_2 = (\overline{w}_1 x) p_1 + (\overline{w}_1 y) q_1 + (\overline{w}_1) r_1 + (\overline{w}_2 x) p_2 + (\overline{w}_2 y) q_2 + (\overline{w}_2) r_2 \tag{11}$$

ANFIS was trained using 70% of total data (randomly selected). Input variables were time-series data. The training parameter was the type of the membership function (MF). Because it has the maximum effect on the accuracy and performance of the ANFIS model. Triangular, Trapezoidal, and G-bell types were selected as the frequently used and popular MF types for comparison purposes in the presence of linear output MF type (for its highest accuracy in comparison with constant type MF). Other parameters like the number of MF types and hybrids method were considered to be constant because they didn't have any significant effect on the modeling procedure in the present study. One of the main reasons can be the dimension of the dataset in the present study. The rest of the data set (30% of the total dataset) was employed for the testing step.

### 3.3. Accuracy Metrics

To compare the predictive power and accuracy performance of MLP and ANFIS two evaluation criteria namely RMSE and determination coefficient (R) are measured for both models. Equations 12 and 13 respectively show how to calculate RMSE and $R^2$.

$$RMSE = \sqrt{\frac{1}{N}\sum_{i=1}^{N}(A - P)^2} \quad (12)$$

$$R^2 = 1 - \left(\frac{\sum_{i=1}^{n}(A-P)^2}{\sum_{i=1}^{n}(A)^2}\right) \quad (13)$$

Where A is the target values and P refers to the predicted values (output of models) and N is the number of data. Using these performance parameters, the accuracy of models can be calculated for comparison purposes.

## 4. Results

In this study, the process of selecting the appropriate model with better predictive power was designed in such a way that the models were first trained by 70% of the data. After the training phase, the predictive performance of the models was tested on the remaining 30% of the data, and then the accuracy of the models was measured and compared by accuracy metrics RMSE and $R^2$. Table 1 shows that the variables of $x_{t-1}$, $x_{t-2}$, and $x_{t-3}$, which are respectively the representation of live animals, animals slaughtered, and livestock yield, are the inputs variables of livestock production quantity and $x_{t-4}$, $x_{t-5}$, which are respectively the representation of yield and losses of agricultural productions, are the inputs variables agricultural productions. In other words, the current model constitutes two outputs: 1) livestock production and 2) agricultural production.

**Table 1.** The prepared dataset for time-series prediction.

| Inputs | Outputs |
|---|---|
| $x_{t-1}$, $x_{t-2}$, $x_{t-3}$, $x_{t-4}$, $x_{t-5}$ | O1= Livestock production |
| | O2= Agricultural production |

### 4.1. Training results

As it is mentioned above, 70% of the data are used to train the models. The training phase was repeated three times, with each model being tested with a different number of neurons.

By changing the number of neurons, the accuracy of the MLP model can be controlled and it reveals the most accurate model. Table 2 shows that in the training phase of the MLP model with the number of neurons ten, fourteen and eighteen were tested. At this stage, the model with ten neurons for predicting livestock production and the model with 18 neurons for predicting agricultural production had the best performance because the corresponding RMSEs were lower compared to other models.

**Table 2.** RMSE results for MLP models with different numbers of neurons in the training phase.

| Variable | Neuron number | RMSE |
|---|---|---|
| Livestock Production | 10 | 275284878.3 |
| Livestock Production | 14 | 462563347.1 |
| Livestock Production | 18 | 320412824.4 |
| Agri. Production | 10 | 36325828 |
| Agri. Production | 14 | 77746693.65 |
| Agri. Production | 18 | 35410107.42 |

On the other hand, to control the accuracy of the ANFIS model in the training phase, the predictive accuracy of different membership functions (MF) was tested. In this study, Triangular-shaped (Tri.), Trapezoidal-shaped (Trap.), and Generalized bell-shaped (Gbell) built-in membership functions are evaluated. The results of the evaluation of the accuracy of MFs are presented in Table 3.

The results show that the model with Trap. built-in membership function has the highest accuracy for predicting both livestock and agricultural production because the RMSE of this model is 4080579.79 for livestock Production and 987950.19 for agricultural production, which are lower than other membership functions. Comparison of Tables 2 and Table 3 illustrates that the performance of the ANFIS model compared to the MLP model in predicting both agricultural and livestock production has been higher. Because the values of RMSE of this model in all cases was lower than the MLP model in the training phase.

**Table 3.** RMSE results for ANFIS models with different MF types in the training phase.

| Variable | MF type | RMSE |
|---|---|---|
| Livestock Production | Tri. | 17225511.04 |
| Livestock Production | Trap. | 4080579.79 |
| Livestock Production | Gbell | 6750734 |
| Agri. Production | Tri. | 2144876.04 |
| Agri. Production | Trap. | 987950.19 |
| Agri. Production | Gbell | 9751562 |

*4.2. Testing results*

After training the models, the models are tested by 30 percent of the data to examine the predictivity power of models. The results of the testing phase of the MLP model are in accordance with the results of the training phase as the MLP model with ten neurons has the highest accuracy for predicting livestock Production because the RMSE of this model is equal to 265590099.2, which is lower than other models with different neurons. In addition, the RMSE of the MLP model 18 neurons for testing agricultural production is 33575595.74 that is lower than the other models indicating the higher accuracy of this model compare to the other models (See Table 4).

Table 4. RMSE results for MLP models with different numbers of neurons in the testing phase.

| Variable | Neuron number | RMSE |
| --- | --- | --- |
| Livestock Production | 10 | 265590099.2 |
| Livestock Production | 14 | 457160675.6 |
| Livestock Production | 18 | 311543277.9 |
| Agri. Production | 10 | 40310186.93 |
| Agri. Production | 14 | 82380698.29 |
| Agri. Production | 18 | 33575595.74 |

However, Table 5 shows that in the testing phase, the ANFIS model with the Gbell membership function had more accurate results with less error levels in both livestock Production prediction (with RMSE=6052851.43) and agricultural Production prediction (with RMSE=1724426) while in the training phase the Trap. membership function model had the highest accuracy rate. Comparing the results of the testing phase with the training phase is the same, and in both phases, the ANFIS model provided higher performance than the MLP model due to the low level of error in predicting both livestock and agricultural production. Therefore, the present study proposes the ANFIS model for predicting food Production.

Table 5. RMSE results for ANFIS models with different MF types in the testing phase.

| Variable | MF type | RMSE |
| --- | --- | --- |
| Livestock Production | Tri. | 11124369 |
| Livestock Production | Trap. | 17894505.8 |
| Livestock Production | Gbell | 6052851.43 |
| Agri. Production | Tri. | 2264668 |
| Agri. Production | Trap. | 2415988 |
| Agri. Production | Gbell | 1724426 |

The coefficient of determination of the ANFIS model was also tested. Figure 4 discloses that the coefficient of determination ($R^2$) of the ANFIS model is very high for both livestock and agricultural production forecast so that $R^2$ is equal to 0.99 for livestock production and 0.94 for agricultural production.

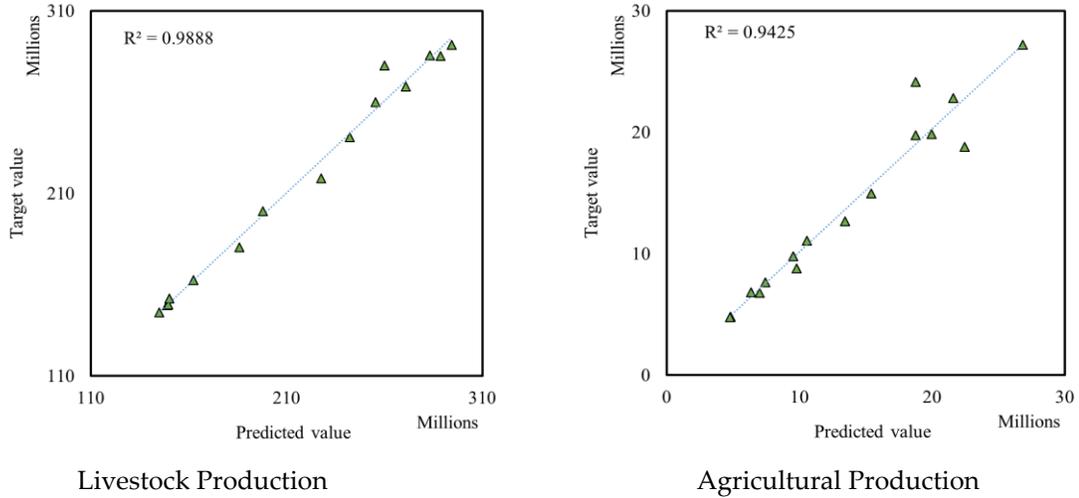

Livestock Production            Agricultural Production

**Figure 4.** The coefficient of determination of the ANFIS model for livestock and agricultural production prediction.

*4.3 Prediction Results*

The results showed that the ANFIS model with Gbell membership function, due to the lower RMSE, not only had a better predictive performance in both agricultural and livestock production forecasting compared to the ANFIS model with other membership functions, but also it has a higher predictability power on the current data compared to the MLP model. Consequently, this model was selected to predict food production in Iran. The results of the prediction of Iranian agricultural and livestock products for 2018 to 2030 using the ANFIS model with Gbell membership function are presented in Table 6.

**Table 6.** Prediction of agricultural and livestock production in Iran

| Year | Gbell Livestock products | Gbell Agricultural products |
| --- | --- | --- |
| 2018 | 351165674 | 30231125 |
| 2019 | 351393213.3 | 30242351 |
| 2020 | 351889340.6 | 30282632 |
| 2021 | 353044979.8 | 30413014 |
| 2022 | 355433096.5 | 30700922 |
| 2023 | 359042959.3 | 31147556 |
| 2024 | 363583520.4 | 31717244 |
| 2025 | 368812282 | 32374680 |
| 2026 | 374726642.9 | 33113468 |
| 2027 | 381123336.2 | 33906210 |
| 2028 | 387439028 | 34691535 |
| 2029 | 393045557.7 | 35395595 |
| 2030 | 397788163.4 | 35992727 |

In order to better represent the predicted trend for agricultural and livestock products in Iran, Figure 5 is designed based on the predicted data. Figure 5 shows that agricultural and livestock products in Iran are expected to have an upward trend with almost the same slope. This is because the predictive model of this study, using time series data, predicts that food production in Iran will increase in the upcoming decade.

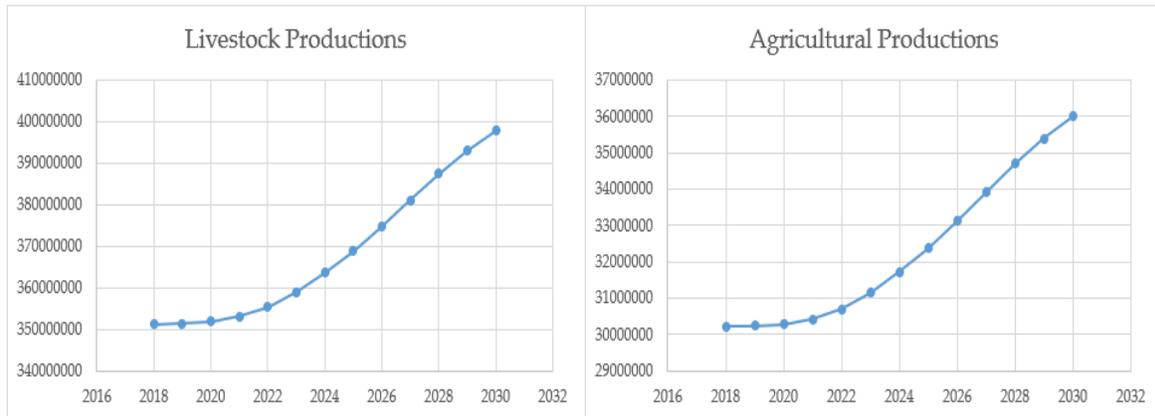

Figure 5. The result of predicting agricultural and livestock production for the next ten years in Iran

5. Conclusions

As the world's population grows, so does the demand for food, and in recent years the number of people exposed to hunger, and even severe hunger, is increasing daily. Governments and organizations active in the food industry are planning and preparing to prevent potential problems that may arise in the way of food security for future generations. To achieve food security goals, food is mainly supplied through domestic production and import. Therefore, studying a country's potential for food supply is the first step in planning for food security. Food production prediction gives a realistic view to policy makers and activists in the agricultural and food industries for long-term and short-term planning. Therefore, the present study tried to provide a suitable model with high predictive performance for predicting food production. The present study predicted Iran's agricultural and livestock production for the next ten years. According to the results, it is predicted that in the next ten years, the volume of both agricultural and livestock production in Iran will increase. The findings of this study provide a basis for planning the production volume required for the coming years, planning for budgeting and agricultural subsidies, planning for the active workforce in the agricultural and livestock sectors. In addition, according to forecasts, decision-makers can plan to import needed food production and export surplus domestic production.

Using machine learning, researchers have come up with creative and precise solutions to a variety of food and agricultural problems, such as crop yields prediction. However, there is no research to predict food production. The present study used machine learning models to predict agricultural and livestock products in Iran. For this purpose, the performance of two models, MLP and ANFIS, was tested using time series data of agricultural and livestock production in Iran. The results of accuracy metrics revealed that the ANFIS model has higher predictive power than the MLP model due to its higher predictive accuracy. The current study contributes to food security research by providing a repayable tool to predict the future of agricultural and livestock production. Researchers and decision-makers can use this model to predict the future of food security in a region. Therefore, for future research, it is suggested that using the proposed model of the present study to predict food production in different countries and provide appropriate solutions to combat food insecurity.

One of the limitations of this study is that forecasts for agricultural and livestock production are based only on time series data while other factors such as climate, government policies, and technological advances are considered constant. Another limitation of this article is the generalization of the finding that the ANFIS model outperforms the MLP model because this finding is limited to the time series data of Iran and the result may differ in data related to another country.

**Author Contributions:** conceptualization, S.N., Z.L.; methodology, A.M., S.A; investigation, S.N.; software, S.A.; formal analysis, A.M., S.A.; writing—original draft preparation, S.N.; writing—review and editing, C.M., Z.L.; visualization, A.M.; supervision, A.M., Z.L.

**Acknowledgments:** Support of the European Commission is acknowledged.

**Conflicts of Interest:** The authors declare no conflict of interest.

**Appendix A**

**Table A1.** Abbreviations

| Acronym | Phrases |
|---------|---------|
| ANFIS | adaptive network-based fuzzy inference system |
| ANN | Artificial neural network |
| CNN | Convolutional neural network |
| DS | Data science |
| ML | Machine learning |
| MLP | Multilayer perceptron |
| SOM | Self-Organising Map |
| SSAE | stacked sparse autoencoder (SSAE) |
| SVM | Support vector machine |

**References**

1. Development, I.F.f.A.; UNICEF; Programme, W.F.; Organization, W.H. The state of food security and nutrition in the World: Safeguarding against economic slowdowns and downturns; FAO: 2019.
2. Wunderlich, S.M.; Martinez, N.M. Conserving natural resources through food loss reduction: Production and consumption stages of the food supply chain. International Soil and Water Conservation Research 2018, 6, 331-339.
3. Nosratabadi, S.; Mosavi, A.; Lakner, Z. Food Supply Chain and Business Model Innovation. Foods 2020, 9, 132.
4. Kang, Y.; Khan, S.; Ma, X. Climate change impacts on crop yield, crop water productivity and food security–A review. Progress in natural Science 2009, 19, 1665-1674.

5. Iizumi, T.; Sakuma, H.; Yokozawa, M.; Luo, J.-J.; Challinor, A.J.; Brown, M.E.; Sakurai, G.; Yamagata, T. Prediction of seasonal climate-induced variations in global food production. Nature climate change 2013, 3, 904-908.
6. Hertel, T.W.; Burke, M.B.; Lobell, D.B. The poverty implications of climate-induced crop yield changes by 2030. Global Environmental Change 2010, 20, 577-585.
7. Nosratabadi, S.; Karoly, S.; Beszedes, B.; Felde, I.; Ardabili, S.; Mosavi, A. Comparative Analysis of ANN-ICA and ANN-GWO for Crop Yield Prediction. In Proceedings of 2020 RIVF International Conference on Computing and Communication Technologies (RIVF), Ho Chi Minh, Vietnam, Vietnam, 14-15 Oct. 2020.
8. Pantazi, X.E.; Moshou, D.; Alexandridis, T.; Whetton, R.L.; Mouazen, A.M. Wheat yield prediction using machine learning and advanced sensing techniques. Computers and Electronics in Agriculture 2016, 121, 57-65.
9. Sengupta, S.; Lee, W.S. Identification and determination of the number of immature green citrus fruit in a canopy under different ambient light conditions. Biosystems Engineering 2014, 117, 51-61.
10. Morales, I.R.; Cebrián, D.R.; Blanco, E.F.; Sierra, A.P. Early warning in egg production curves from commercial hens: A SVM approach. Computers and Electronics in Agriculture 2016, 121, 169-179.
11. Alonso, J.; Villa, A.; Bahamonde, A. Improved estimation of bovine weight trajectories using Support Vector Machine Classification. Computers and electronics in agriculture 2015, 110, 36-41.
12. Alonso, J.; Castañón, Á.R.; Bahamonde, A. Support Vector Regression to predict carcass weight in beef cattle in advance of the slaughter. Computers and electronics in agriculture 2013, 91, 116-120.
13. Karandish, F.; Salari, S.; Darzi-Naftchali, A. Application of virtual water trade to evaluate cropping pattern in arid regions. Water resources management 2015, 29, 4061-4074.
14. Qasemipour, E.; Abbasi, A. Virtual water flow and water footprint assessment of an arid region: A case study of South Khorasan province, Iran. Water 2019, 11, 1755.
15. Paymard, P.; Yaghoubi, F.; Nouri, M.; Bannayan, M. Projecting climate change impacts on rainfed wheat yield, water demand, and water use efficiency in northeast Iran. Theoretical and Applied Climatology 2019, 138, 1361-1373.
16. Raeisi, L.G.; Morid, S.; Delavar, M.; Srinivasan, R. Effect and side-effect assessment of different agricultural water saving measures in an integrated framework. Agricultural Water Management 2019, 223, 105685.
17. Akhoundi, A.; Nazif, S. Sustainability assessment of wastewater reuse alternatives using the evidential reasoning approach. Journal of cleaner production 2018, 195, 1350-1376.
18. Mehri, N.; Messkoub, M.; Kunkel, S. Trends, determinants and the implications of population aging in Iran. Ageing International 2020, 1-17.
19. Ekhlaspour, P.; Foroumandi, E.; Ebrahimi-Mameghani, M.; Jafari-Koshki, T.; Arefhosseini, S.R. Household food security status and its associated factors in Baft-Kerman, IRAN: a cross-sectional study. Ecology of Food and Nutrition 2019, 58, 608-619.
20. Esfarjani, F.; Hosseini, H.; Khaksar, R.; Roustaee, R.; Alikhanian, H.; Khalafi, M.; Khaneghah, A.M.; Mohammadi-Nasrabadi, F. Home Food Safety Practice and Household Food Insecurity: A Structural Equation Modeling Approach. Iranian Journal of Public Health 2019, 48, 1870.
21. Fathi Beyranvand, H.; Eghtesadi, S.; Atai Jafari, A.; Movahedi, A. Prevalence of Food Insecurity in Pregnant Women in Khorramabad City and its Association with General Health and other Factors. Iranian Journal of Nutrition Sciences & Food Technology 2019, 14, 21-30.
22. Alamdarlo, H.N.; Riyahi, F.; Vakilpoor, M.H. Wheat self-sufficiency, water restriction and virtual water trade in Iran. Networks and Spatial Economics 2019, 19, 503-520.
23. Tealab, A. Time series forecasting using artificial neural networks methodologies: A systematic review. Future Computing and Informatics Journal 2018, 3, 334-340.
24. Tealab, A.; Hefny, H.; Badr, A. Forecasting of non-linear time series using ANN. Future Computing and Informatics Journal 2017, 2, 39-47.
25. Karandish, F. and A. Hoekstra. Informing national food and water security policy through water footprint assessment: the case of Iran. Water, 2017, 9, 831.
26. Esfahani, A.K., et al. Overseas cultivation: the complimentary approach for developing food security. Bulgarian Journal of Agricultural Science, 2019, 25, 26-35.
27. Vargas, R.; Mosavi, A.; Ruiz, R. Advances in Intelligent Systems and Computing, 2017, 7, 122–148.
28. Sengupta, S.; Lee,W.S. Identification and determination of the number of immature green citrus fruit in a canopy under different ambient light conditions. Biosystems Engineering, 2014, 117, 51–61.
29. Su, Y.; Xu, H.; Yan, L. Support vector machine-based open crop model (SBOCM): Case of rice production in China. Saudi Journal of Biological Sciences, 2017, 24, 537–547.

30. Ali, I.; Cawkwell, F.; Dwyer, E.; Green, S. Modeling Managed Grassland Biomass Estimation by Using Multitemporal Remote Sensing Data—A Machine Learning Approach. IEEE Journal of Selected Topics in Applied Earth Observations and Remote Sensing, 2016, 10, 3254–3264.
31. Chung, C.L.; Huang, K.J.; Chen, S.Y.; Lai, M.H.; Chen, Y.C.; Kuo, Y.F. Detecting Bakanae disease in rice seedlings by machine vision. Computers and Electronics In Agriculture, 2016, 121, 404–411.
32. Ebrahimi, M.A.; Khoshtaghaza, M.H.; Minaei, S.; Jamshidi, B. Vision-based pest detection based on SVM classification method. Comput. Computers and Electronics in Agriculture, 2017, 137, 52–58.
33. Moshou, D.; Bravo, C.; West, J.; Wahlen, S.; McCartney, A.; Ramon, H. Automatic detection of "yellow rust" in wheat using reflectance measurements and neural networks. Computers and electronics in agriculture, 2004, 44, 173–188.
34. Moshou, D.; Bravo, C.; Oberti, R.; West, J.; Bodria, L.; McCartney, A.; Ramon, H. Plant disease detection based on data fusion of hyper-spectral and multi-spectral fluorescence imaging using Kohonen maps, Real-Time Imaging 2005, 11, 75–83.
35. Pantazi, X.E.; Tamouridou, A.A.; Alexandridis, T.K.; Lagopodi, A.L.; Kashefi, J.; Moshou, D. Evaluation of hierarchical self-organising maps for weed mapping using UAS multispectral imagery. Computers and Electronics in Agriculture, 2017, 139, 224–230.
36. Pantazi, X.-E.; Moshou, D.; Bravo, C. Active learning system for weed species recognition based on hyperspectral sensing. Biosystems Engineering, 2016, 146, 193–202.
37. Feng, Y.; Peng, Y.; Cui, N.; Gong, D.; Zhang, K. Modeling reference evapotranspiration using extreme learning machine and generalized regression neural network only with temperature data. Computers and Electronics in Agriculture, 2017, 136, 71–78.
38. Patil, A.P.; Deka, P.C. An extreme learning machine approach for modeling evapotranspiration using extrinsic inputs. Computers and Electronics in Agriculture, 2016, 121, 385–392.
39. Craninx, M.; Fievez, V.; Vlaeminck, B.; De Baets, B. Artificial neural network models of the rumen fermentation pattern in dairy cattle. Computers and Electronics in Agriculture, 2008, 60, 226–238.
40. Alonso, J.; Villa, A.; Bahamonde, A. Improved estimation of bovine weight trajectories using Support Vector Machine Classification. Computers and Electronics in Agriculture, 2015, 110, 36–41.
41. Alonso, J.; Castañón, Á.R.; Bahamonde, A. Support Vector Regression to predict carcass weight in beef cattle in advance of the slaughter. Computers and Electronics in Agriculture, 2013, 91, 116–120.
42. Liu, Z., He, Y., Cen, H., & Lu, R. Deep feature representation with stacked sparse auto-encoder and convolutional neural network for hyperspectral imaging-based detection of cucumber defects. Transactions of the ASABE, 2018, 61, 425–436. https://doi.org/10.13031/trans.12214
43. Rodriguez, F. J., Garcia, A., Pardo, P. J., Chavez, F., & Luque-Baena, R. M. Study and classification of plum varieties using image analysis and deep learning techniques. Progress in Artificial Intelligence, 2018, 7, 119–127. https://doi.org/10.1007/s13748-017-0137-1
44. Azizah, L. M., Umayah, S. F., Riyadi, S., Damarjati, C., & Utama, N. A. Deep learning implementation using convolutional neural network in mangosteen surface defect detection. In 7th IEEE International Conference on Control System, Computing and Engineering (ICCSCE), 2017, 242–246. https://doi.org/10.1109/ICCSCE.2017.8284412
45. Cheng, J. H., & Sun, D. W. Partial least squares regression (PLSR) applied to NIR and HSI spectral data modeling to predict chemical properties of fish muscle. Food Engineering Reviews, 2017, 9, 36–49. https://doi.org/10.1007/s12393-016-9147-1
46. Yu, X. J., Wang, J. P., Wen, S. T., Yang, J. Q., & Zhang, F. F. A deep learning based feature extraction method on hyperspectral images for nondestructive prediction of TVB-N content in Pacific white shrimp (Litopenaeus vannamei). Biosystems Engineering, 2019, 178, 244–255. https://doi.org/10.1016/j.biosystemseng.2018.11.018
47. Bisgin, H., Bera, T., Ding, H. J., Semey, H. G., Wu, L. H., . . . Xu, J. Comparing SVM and ANN based machine learning methods for species identification of food contaminating beetles. Scientific Reports, 2018, 8, 12. https://doi.org/10.1038/s41598-018-24926-7
48. Song, Q., Zheng, Y. J., Xue, Y., Sheng, W. G., & Zhao, M. R. An evolutionary deep neural network for predicting morbidity of gastrointestinal infections by food contamination. Neurocomputing 2017, 226, 16–22. https://doi.org/10.1016/j.neucom.2016.11.018
49. Boniecki. P, Zaborowicz. M, Pilarska. A, Piekarska-Boniecka. H. Identification Process of Selected Graphic Features Apple Tree Pests by Neural Models Type MLP, RBF and DNN. *Agriculture*. 2020, *10*, 218.
50. Olier. I, Sansom. A, Lisboa. P, Ortega-Martorell. S. Using MLP partial responses to explain in-hospital mortality in ICU. *IEEE International Conference on Data Analytics for Business and Industry: Way Towards a Sustainable Economy (ICDABI)*, 2020, 1-5.

51. Ying, L.-C.; Pan, M.-C. Using adaptive network based fuzzy inference system to forecast regional electricity loads. Energy Conversion and Management 2008, 49, 205-211.